\documentclass[reprint, superscriptaddress, amsmath, amssymb, aps]{revtex4-2}

\usepackage{bm}
\usepackage[pdftex]{graphicx}
\usepackage{amsmath}
\usepackage{amssymb}
\usepackage{physics}
\usepackage{mathcomp}
\usepackage{ascmac}
\usepackage{enumerate}
\usepackage{url}
\usepackage{dcolumn}
\usepackage{float}

\begin{document}


\title{Propagation and circulating modes of reciprocal non-Hermitian skin effect}


\author{Issei Takeda}
\affiliation{Department of physics, Institute of Science Tokyo, 2-12-1 Ookayama, Meguro-ku, Tokyo 152-8550, Japan}
\affiliation{NTT Basic Research Laboratories, NTT Corporation, 3-1 Morinosato Wakamiya, Atsugi-shi, Kanagawa 243-0198, Japan}
\author{Taiki Yoda}
\affiliation{Department of physics, Institute of Science Tokyo, 2-12-1 Ookayama, Meguro-ku, Tokyo 152-8550, Japan}
\author{Yuto Moritake}
\affiliation{Department of physics, Institute of Science Tokyo, 2-12-1 Ookayama, Meguro-ku, Tokyo 152-8550, Japan}
\author{Kenta Takata}
\affiliation{NTT Basic Research Laboratories, NTT Corporation, 3-1 Morinosato Wakamiya, Atsugi-shi, Kanagawa 243-0198, Japan}
\affiliation{Nanophotonics Center, NTT Corporation, 3-1 Morinosato Wakamiya, Atsugi-shi, Kanagawa 243-0198, Japan}

\author{Masaya Notomi}
\affiliation{Department of physics, Institute of Science Tokyo, 2-12-1 Ookayama, Meguro-ku, Tokyo 152-8550, Japan}
\affiliation{NTT Basic Research Laboratories, NTT Corporation, 3-1 Morinosato Wakamiya, Atsugi-shi, Kanagawa 243-0198, Japan}
\affiliation{Nanophotonics Center, NTT Corporation, 3-1 Morinosato Wakamiya, Atsugi-shi, Kanagawa 243-0198, Japan}

\date{\today}

\begin{abstract}

  The non-Hermitian skin effect (NHSE) is a novel localization phenomenon, in which all bulk states 
  in a non-Hermitian system under certain conditions are localized at the edge of the system. 
  Conventionally, most studies of NHSE have dealt with discrete lattice systems with non-reciprocal 
  couplings. However in recent years, NHSE in a reciprocal two-dimensional continuous medium, 
  such as photonic crystal systems, has also been reported. In particular, we have previously 
  shown that NHSE also occurs in two-dimensional uniform media. 
  In such two-dimensional systems, skin modes propagate in a direction perpendicular to the localization direction, 
  and especially, they have the property of propagating in only one direction.
  In this paper, we show numerically an intriguing scattering phenomenon: 
  when a scatterer is placed in the path of a skin mode, the scattering causes the skin mode to hop between opposing edges.
  In addition, we propose a new method of generating circulating modes with orbital angular momentum using 
  this scattering phenomenon. Our work paves the way for new applications of NHSE as micro-sized 
  optical devices manipulating or generating OAM.

\end{abstract}


\maketitle

\section{Introduction}
Non-Hermitian systems with gain or loss have attracted much attention in recent years 
because of the appearance of novel phenomena that have not been realized in Hermitian systems. 
Optical systems are a suitable platform for exploring non-Hermitian physics because
non-Hermiticity can be easily introduced by gains and losses caused by materials. 
In particular, studies in optical systems with parity-time ($\mathcal{PT}$) symmetry \cite{PT} have been intensive, 
and many peculiar phenomena have been reported, such as symmetric light propagation 
\cite{reci1, reci2, reci3}, unidirectional invisibility \cite{invi1,invi2,invi3}, and fast-light states \cite{fast}.

\begin{figure}[t]
  \includegraphics[width=0.75\columnwidth]{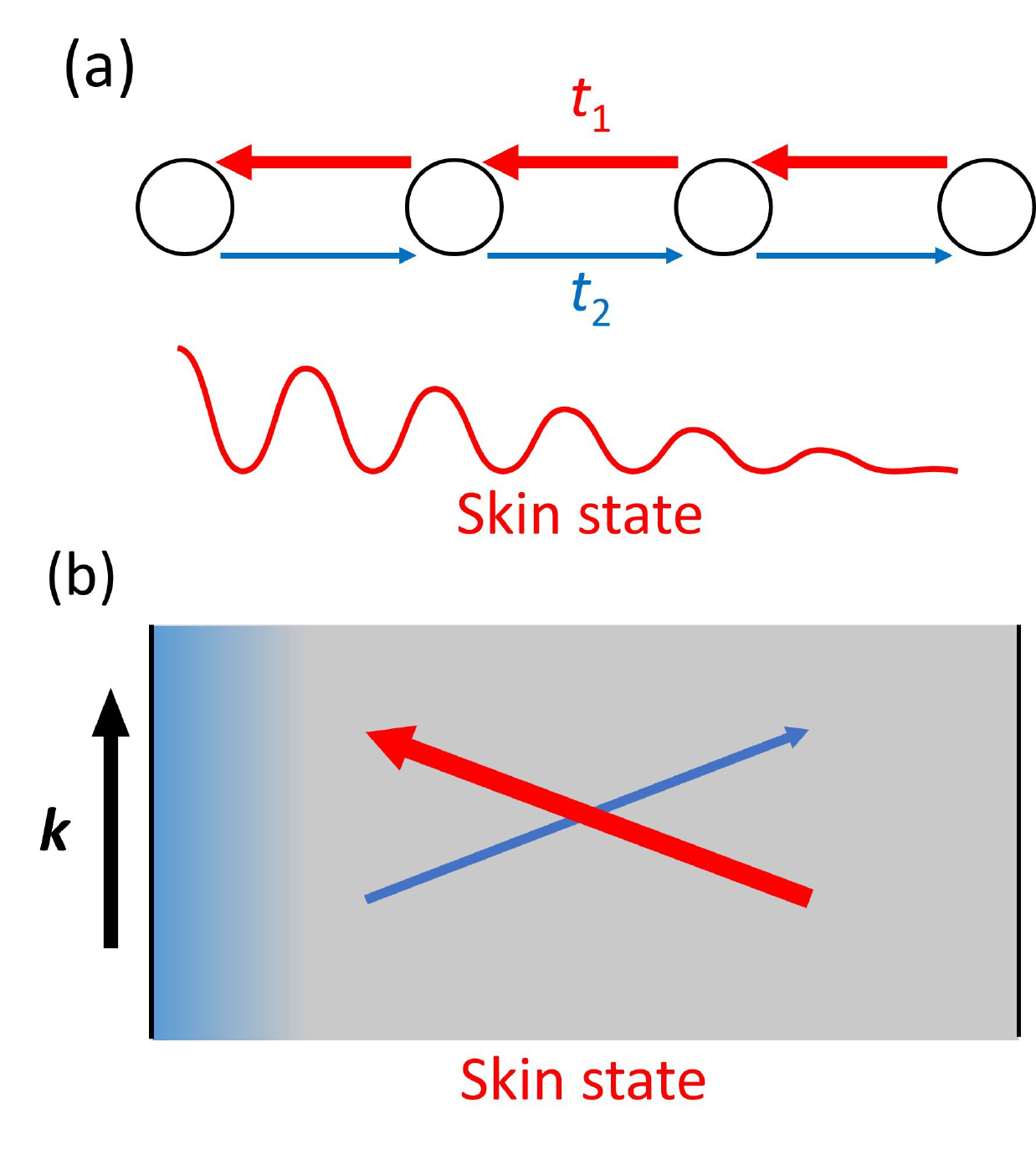}
  \caption{(a) Skin state in 1-dimensional crystal. (b) Skin state in 2-dimensional continuous medium. 
  (c)(d)Schematics of anisotropic two-dimensional uniform medium and $\theta$-tilted 
  anisotropic two-dimensional uniform medium, respectively.}\label{intro}
\end{figure}

One of the phenomena unique to non-Hermitian systems is the non-Hermitian skin effect (NHSE) \cite{skin-exam1,skin-exam2,skin-exam3,skin-exam4,skin-exam5}, 
a phenomenon in which all bulk eigenstates are localized at the edge of the system under open boundary 
conditions in certain non-Hermitian systems. In Hermitian systems and non-Hermitian systems without NHSE,
bulk states extended throughout the system usually form a continuous energy spectrum. On the other hand, in non-Hermitian 
systems where NHSE occurs, localized states at the edges of the system form a continuous spectrum.
NHSE was first reported in a one-dimensional tight binding model with asymmetric coupling rates as shown in Fig. \ref{intro}(a) \cite{skin-exam1}. Since then, studies 
of NHSE have focused on discrete lattice systems described by the tight binding model, and experiments have been reported 
on various systems described by the tight binding model \cite{skin-exp1,skin-exp2,skin-exp3,skin-exp4,skin-exp5,skin-exp6,skin-exp7,skin-exp8,skin-exp9,skin-exp10,skin-exp11}.

However, in recent years, NHSE has also been reported in continuous systems 
such as anisotropic photonic crystals \cite{phc1,phc2,phc3,phc4,phc5,phc6,phc7,phc8} and 
anisotropic uniform media \cite{yoda}, which are described by differential 
equations such as Maxwell's equations. In particular, propagation has not been considered 
in the conventional NHSE studies. However, in a continuous medium, skin modes become two-dimensional 
propagating states because they have a wavevector component in the direction perpendicular 
to the localization direction (Fig. \ref{intro}(b)). 
Previous studies of NHSE in optical systems have shown that in infinitely long stripe systems, 
i.e., with periodic boundary conditions in one direction, the localized position of skin modes depends 
on the propagation direction, and when the propagation direction is reversed, 
the localized position also reverses \cite{phc1,phc3,yoda}. Therefore, skin modes propagate in only one 
direction at one edge. 
However, there are no reports on the properties of the propagating skin modes, such as transmission or reflection.
Moreover, there have been some studies of NHSE in finite systems \cite{phc4,phc6,phc7}, 
but none of them have focused on unidirectional propagation.


In this work, we numerically investigate properties of skin mode propagation 
in finite two-dimensional systems. First, we simulate the skin mode 
in the system with the mirror-time ($\mathcal{MT}$) symmetric medium \cite{yoda,RT}, an anisotropic medium 
with balanced gain and loss and a scatterer in the path of the skin mode. 
Then, we found a intriguing scattering phenomenon in which the complete backscattering 
of the skin mode is suppressed, but instead hopping to the opposite edge. 
Also, the excited skin mode and the hopped skin mode propagated in opposite 
directions, and in total, we confirmed that it was a rotation-like mode.

Furthermore, 
we have applied the scattering phenomenon
of skin modes to consider closed systems surrounded by reflective boundaries.
Initially, we considered a square structure, and skin modes appeared at the top, 
bottom, left, and right edges. Skin modes on the left and right edges propagated 
clockwise, while skin modes on the top and bottom edges propagated counterclockwise, 
and they were balanced. However, by considering the rectangular structure, one of 
the rotations was strongly manifested, and we succeeded in generating 
the optical orbital angular momentum (OAM) \cite{OAM} by a new method using NHSE.
We also discuss the application of the geometry dependence of NHSE \cite{phc4,phc7,uni,geo1} to improve the efficiency of the OAM.

Finally, we confirmed that OAM generation by NHSE was feasible, even when using loss-bias media, which are easy to realize.

\section{NHSE in uniform media}\label{sec2}
\subsection{Analytical solution of NHSE in uniform media}\label{sec2-a}
Conventional studies of optical NHSE have dealt with systems with a periodically 
modulated dielectric constant, such as photonic crystals. On the other hand, 
we have recently shown that NHSE can be realized even in media with a uniform 
dielectric permittivity \cite{yoda}.
The realization of NHSE requires the non-Hermiticity and the anisotropy of the dielectric 
tensor, and in a uniform medium, it is easy to control them. In addition, the analytical 
solution of NHSE can be derived. Thus, in this work, we deal with the propagation 
of NHSE in a reciprocal uniform medium.

\begin{figure}[t]
  \includegraphics[width=1\columnwidth]{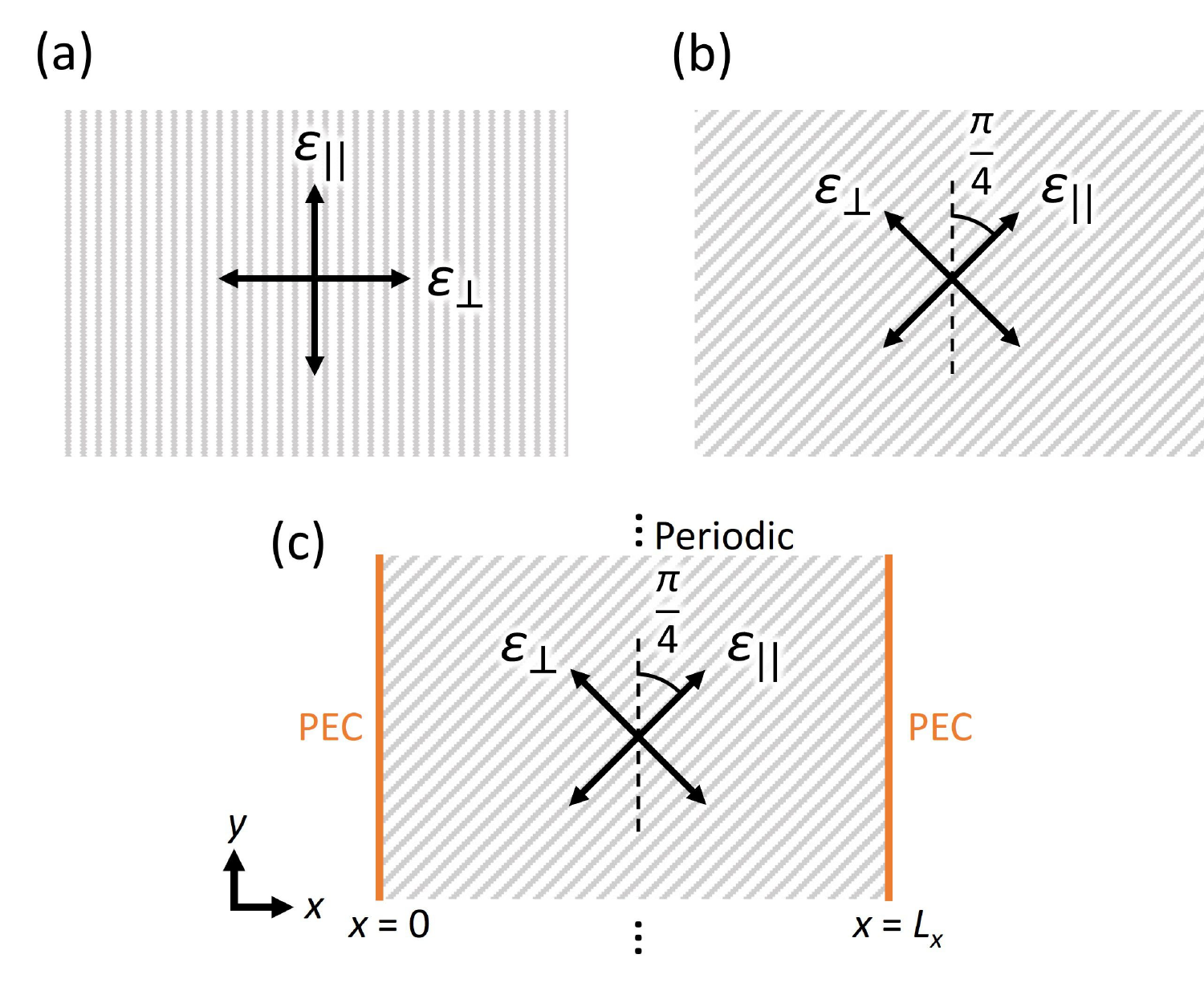}
  \caption{(a)(b) Schematics of anisotropic uniform medium and 45 degrees tilted 
  anisotropic uniform medium, respectively.(c)Infinitely long stripe structure with 45 degrees tilted medium.}\label{aniso}
\end{figure}

First, the dielectric tensor of a medium with non-Hermitian anisotropy, 
as shown in Fig. \ref{aniso}(a), is given by
\begin{equation}
  \mqty(\varepsilon_{\perp} & 0 & 0\\
  0 & \varepsilon_\parallel & 0\\
  0 & 0 & \varepsilon_z),
\end{equation}
where $\varepsilon_{\parallel}$ and $\varepsilon_{\parallel}$ have non-zero imaginary parts.
Here we consider a medium tilted at 45 degrees to this medium 
as shown in Fig. \ref{aniso}(b), 
tilted 45 degrees from this medium. In this case, the dielectric tensor is given by
\begin{equation}
  \mqty(\varepsilon_{xx} & \varepsilon_{xy} & 0\\
  \varepsilon_{yx} & \varepsilon_{yy} & 0\\
  0 & 0 & \varepsilon_{zz})
  =\mqty((\varepsilon_\parallel+\varepsilon_\perp)/2 &(\varepsilon_\parallel-\varepsilon_\perp)/2 & 0\\
  (\varepsilon_\parallel-\varepsilon_\perp)/2 & (\varepsilon_\parallel+\varepsilon_\perp)/2 & 0\\
  0 & 0 & \varepsilon_z ).
\end{equation}
When we consider the transverse electric mode $H_z(x,y,t)=e^{i(\omega t-k_y y)}H_z(x)$ traveling in the $y$ direction 
in a system with finite width $L_x$ with periodic boundary conditions (PBC) imposed in the $y$ direction 
and perfect electric conductors (PEC) boundary condition in the $x$ direction, as shown in Fig. \ref{aniso}(c).
In this system, the eigen states can be obtained analytically \cite{yoda}:

\begin{equation}
  \qty(\frac{\omega}{c})^2=\eta_{yy}\qty(\frac{\pi n}{L_x})^2+(\eta_{xx}-q^2\eta_{yy})k_y^2\ \ \ n=1,2,\ldots,\label{f-k}
\end{equation}
\begin{eqnarray}
  E_y(x)&=&e^{iqk_yx}\sin\qty(\frac{\pi n}{L_x}x),\label{Ey}\\
  H_z(x)&=&\frac{1}{2}ie^{iqk_yx}\Biggl[\qty(\frac{1}{Z_{x+}}-\frac{1}{Z_{x-}})\cos\qty(\frac{\pi n}{L_x}x)\nonumber\\
  &-&  i\qty(\frac{1}{Z_{x+}}+\frac{1}{Z_{x-}})\sin\qty(\frac{\pi n}{L_x}x)\Biggr],\\
  E_x(x)&=&-\frac{1}{2}ie^{iqk_yx}\Biggl[\qty(\frac{Z_{y+}}{Z_{x+}}-\frac{Z_{y-}}{Z_{x-}})\cos\qty(\frac{\pi n}{L_x}x)\nonumber\\
  &-&  i\qty(\frac{Z_{y+}}{Z_{x+}}+\frac{Z_{y-}}{Z_{x-}})\sin\qty(\frac{\pi n}{L_x}x)\Biggr]\label{Ex},
\end{eqnarray}
where $c$ is the speed of light in vacuum, $\eta_{ij}$is the component of the inverse tensor of the dielectric tensor, 
and $q, k_\pm, Z_{x\pm},Z_{y\pm}$ are defined as follows: 
\begin{eqnarray}
  q&=&\frac{\varepsilon_{xy}+\varepsilon_{yx}}{2\varepsilon_{xx}},\label{q}\\
  k_\pm&=&\pm\frac{\pi n}{L_x}-qk_y,\\
  Z_{x\pm}&=&\frac{1}{\omega\varepsilon_0}\qty(-k_y\eta_{yx}+k_\pm\eta_{yy}),\\
  Z_{y\pm}&=&\frac{1}{\omega\varepsilon_0}\qty(k_y\eta_{xx}-k_\pm\eta_{xy}),\label{eq1}
\end{eqnarray}
where $\varepsilon_0$ is the dielectric permittivity in vacuum.
From Eqs. (\ref{Ey})-(\ref{Ex}), the state is localized when $\mathrm{Im}(q)\neq 0$ and $k_y\neq0$.
When the medium is not tilted, $q=0$ and no localization occurs, 
but when tilted, if $\varepsilon_\perp$ and $\varepsilon_\parallel$ are appropriately chosen, 
a nonzero imaginary part of $q$ can be realized and skin modes appear.

Here, from Eqs. (\ref{Ey})-(\ref{Ex}), it can be seen that when the direction 
of skin mode propagation is reversed, i.e., the sign of $k_y$ is reversed, the 
localization position is also reversed. Thus, the skin mode propagates in only 
one direction at one edge.
Similar to the unidirectional propagation seen in topological edge modes \cite{topo1,topo2,topo3}, 
backscattering may be suppressed also in the case of propagating skin modes.

\subsection{$\mathcal{MT}$ symmetric media}\label{MT-setsumei}
We have simulated the propagation of skin modes with $\mathcal{MT}$ symmetric media \cite{yoda,RT}. 
Here, $\mathcal{MT}$ symmetry is the symmetry about mirror and time reversal. 
The mirror plane is a plane perpendicular to the $x$- or $y$-direction. 
The time reversal corresponds to a complex conjugate 
operation of dielectric constants. 
In non-Hermitian systems, $\mathcal{PT}$ symmetry is often discussed. 
$\mathcal{PT}$ and $\mathcal{MT}$ symmetry coincide in one-dimensional systems, 
but in two-dimensional systems, the operation by $\mathcal{P}$ is $(x,y)\mapsto (-x,-y)$, 
while the operation by $\mathcal{M}$ is $(x,y)\mapsto (-x,y)$ or $(x,y)\mapsto (x,-y)$, 
and they are different.

\begin{figure}[t]
  \includegraphics[width=1\columnwidth]{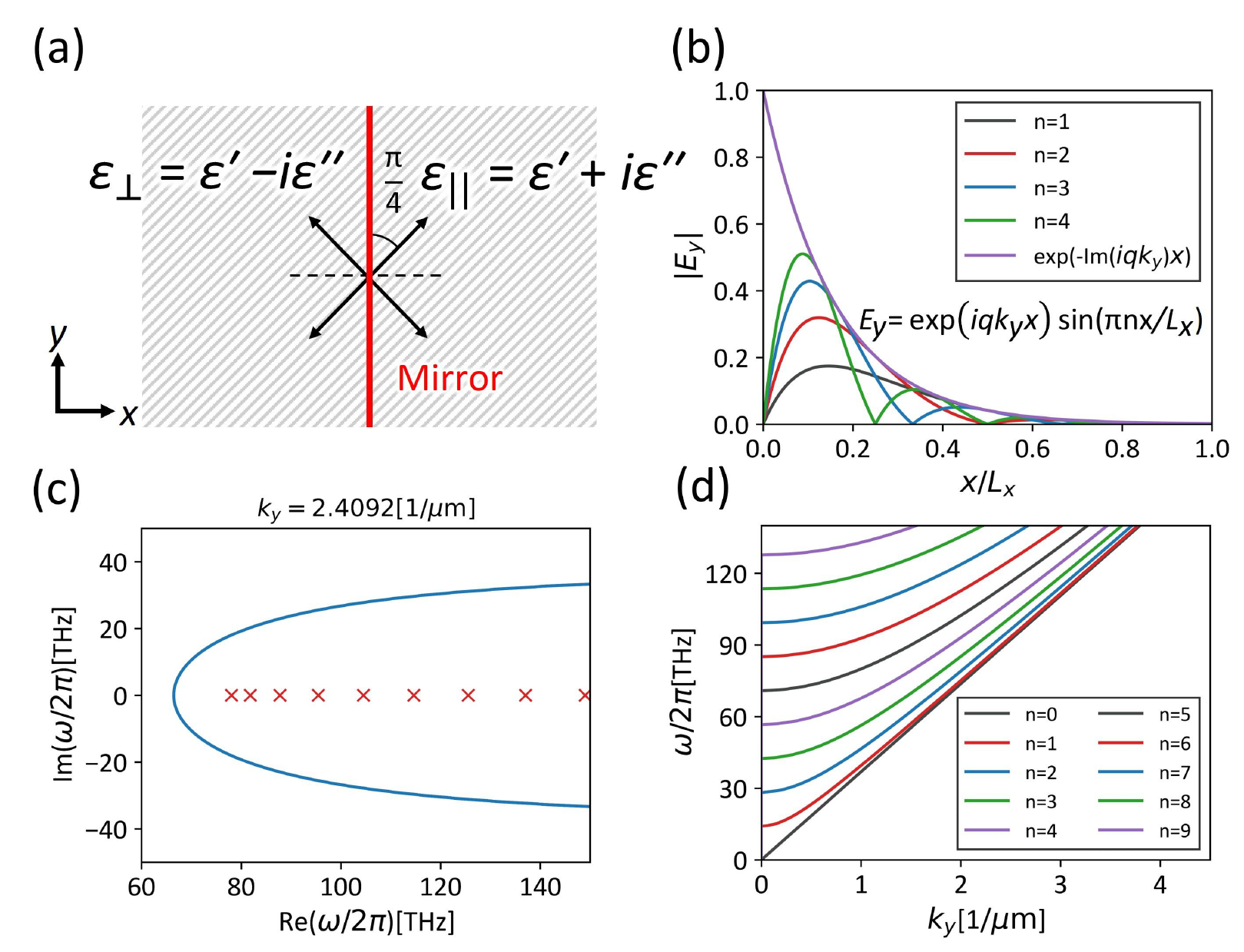}
  \caption{(a) $\mathcal{MT}$ symmetric medium. (b) eigenfrequency spectra under periodic boundary condition 
  and open boundary condition. (c) Relationship between $ky$ and eigenfrequencies in the system shown in (a).
  (d) Eigenmodes in the infinitely long stripe system with $\mathcal{MT}$ symmetric medium.}\label{MTskin}
\end{figure}

As shown in Fig. \ref{MTskin}(a), $\mathcal{MT}$ symmetry is satisfied 
when it holds that $\varepsilon_\parallel=\varepsilon_\perp^*$.
In $\mathcal{MT}$ symmetric media, $q$ is a pure imaginary number. Thus, when $\mathrm{Re}(k_y) \neq 0$, 
eigenstates (Eqs. (\ref{Ey})-(\ref{Ex})) become localized states (Fig. \ref{MTskin}(b)). In a periodic structure like the photonic crystal system, the 
eigenfrequencies of skin modes are distributed in the closed loop of eigenfrequencies 
under the PBC on the complex frequency plane. On the other hand, in a uniform medium, 
the eigenfrequencies under the PBC form an open arc,  and the 
eigenfrequencies of the skin modes are distributed inside the open arc. In particular, 
in $\mathcal{MT}$ symmetric uniform medium, the open arc is symmetric to the real axis 
of the complex frequency plane, and the eigenfrequencies of the skin mode appear 
on the real axis inside the open arc (Fig. \ref{MTskin}(c)). In this case, from Eq. (\ref{f-k}), $k_y$ is also a real number. 
Here, the relationship between $k_y$ and $\omega/2\pi$ is shown in Fig. \ref{MTskin}(d).
This property, where $k_y$ and $\omega$ are both real numbers, 
allows for skin modes that are not amplified or attenuated in space or time, which is ideal for exploring propagation.

Here anisotropy as shown in Fig. \ref{aniso} can be approximately realized by considering the effective medium model for a subwavelength multilayer metamaterial
consisting of two different media, and obtaining $\varepsilon_{\parallel}$ and $\varepsilon_{\perp}$ 
as the dielectric constants in the direction parallel and perpendicular to the multilayer, 
respectively. In particular, the realization of an $\mathrm{MT}$-symmetric medium requires 
both a loss and a gain medium (see Appendix). In this work, we mainly use an $\mathcal{MT}$ symmetric medium ($\varepsilon_\parallel=\varepsilon_\perp^*=1.673+0.96593i$), 
which is obtained from a loss medium and a gain medium with dielectric constants $\varepsilon_1=0.70711-0.70711i$ and 
$\varepsilon_2=2.639+2.639i$, respectively.
In this case, components of the dielectric tensor are $\varepsilon_{xx}=\varepsilon_{yy}=1.673$ and $\varepsilon_{xy}=\varepsilon_{yx}=0.96593i$.

\section{Propagation of skin modes}
\begin{figure*}[ht]
  \includegraphics[width=1.8\columnwidth]{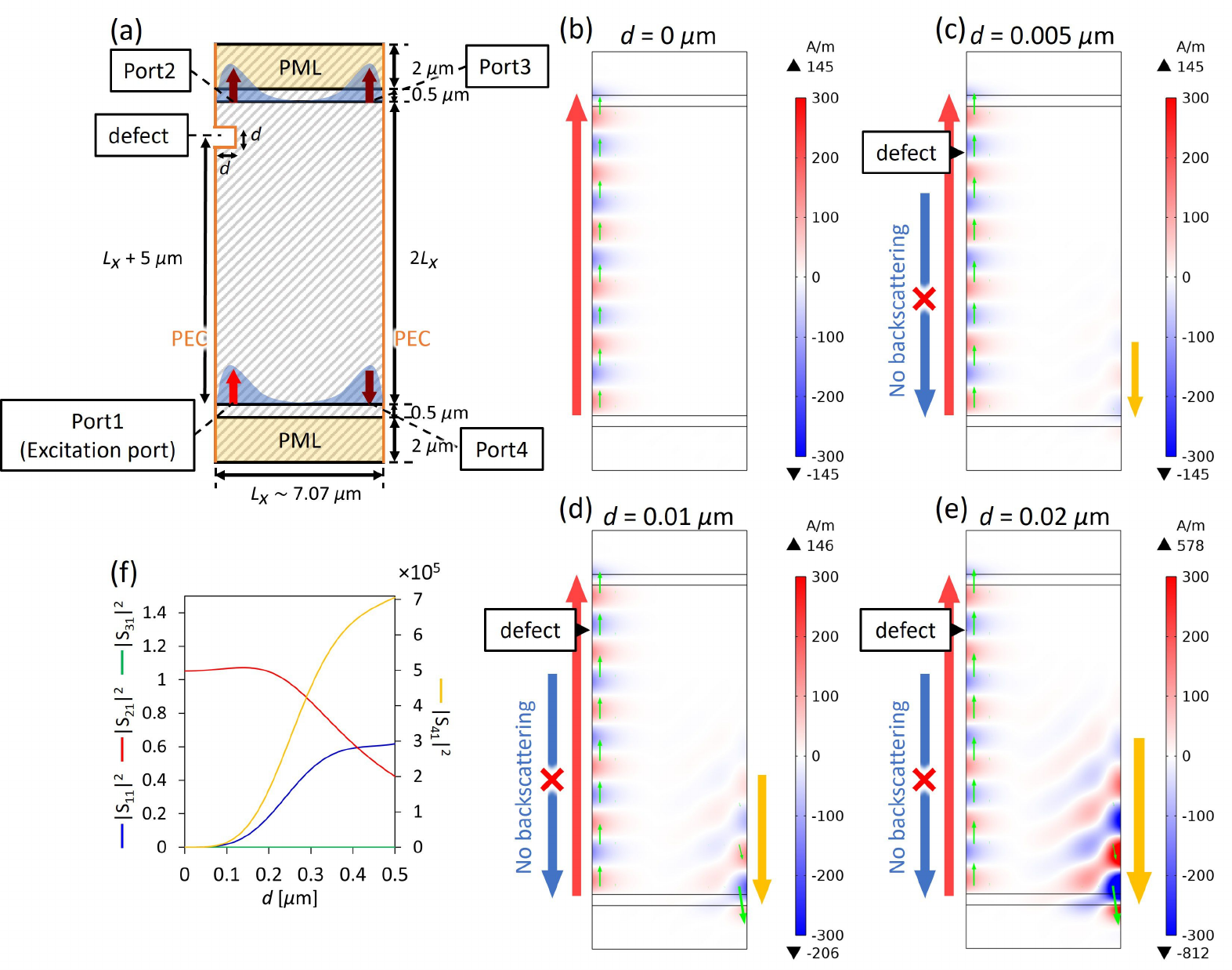}
  \caption{(a) Schematic of the structure of the simulation. (b)(c)(d)(e) Calculated $H_z$ distribution for the length $d$ 
  of one side of the defect for $d=0\mu\mathrm{m}$ (i.e. no defect), $d=0.005\mu\mathrm{m}$, $d=0.01\mu\mathrm{m}$, and $d=0.02\mu\mathrm{m}$, respectively.
  Green arrows show Poynting vectors. Note that the Poynting vectors of the right edge are normalized to be as large as those of the left edge.
  (f) Relationship between transmittance, reflectance and defect edge length $d$.}\label{propa}
\end{figure*}

\subsection{Model for numerical calculation}
A schematic of the structure is shown in Fig. \ref{propa}(a). 
We consider the $\mathcal{MT}$ symmetric medium described above of finite width $L_x\sim7.07\mu\mathrm{m}$ (see Appendix A for how to determine the size of the structure)
sandwiched between PECs in the $x$ direction.
Then, we calculate by the finite element method by using COMSOL Multiphysics.
We have also placed a square defect (length of one side $d$) at the left edge, and a perfectly matched layer (PML) \cite{PML}
at the top and bottom to suppress reflections. 
We excite the propagating skin mode with a finite $k_y$ value at the Port 1 where we set the light field profile for the $n=1$ 
skin mode localized at the left edge determined by Eqs. (\ref{Ey})-(\ref{Ex}).
In addition, Port2-4 are also added for the detection of transmissions and reflections, and 
the scattering parameters ($S_{11}, S_{21}, S_{31}$, and $S_{41}$) 
are calculated for each port. 
Here, Port 2 detects a left localized mode that propagates in the $+y$ direction, same as the mode set for Port 1. 
Also, Port 3 detects the right localized mode that propagates in the $+y$ direction, which is the left-right reversal 
of the mode set at Port 1. Moreover, Port 4 detects the right localized mode that propagates in the $-y$ direction.
The characteristics of the localization mode detected at each detection port are shown in Fig. \ref{propa}(a).

\subsection{Nature of unidirectional propagation}
First, the simulation result exited at 90 THz for the no defect case ($d=0\ \mu \mathrm{m}$) is shown in Fig. \ref{propa} (b). 
Here, $k_y = 2.4092\ 1/\mu\mathrm{m}$ from Eq. (\ref{f-k}).
In this case, skin modes without amplification or attenuation propagating in the $+y$ direction 
appeared. This skin mode should only be allowed to exist in the left edge and propagate in the $+y$ direction due to 
the propagation direction dependence of the localization position. 

Next, Figures \ref{propa}(c)-(e) shows the calculated $H_z$ distributions in a structure with a defect
in the path of the skin mode under the same excitation condition. Figure \ref{propa}(c) shows that the left-localized skin mode is excited 
and looks almost the same as in Fig. \ref{propa}(b). As expected there is no indication of the back reflection at the left edge (Port 1). 
However, there is a finite intensity in the lower part of the right edge (Port 4). This is due to inter-edge scattering caused by the introduction of the defect. 
Here, what is interesting is that corresponding to the fact that left-localized modes are not allowed 
to propagate in the $-y$ direction as eigenmodes in an infinitely long stripe system, 
reflections propagating on the left edge do not appear, 
and instead, inter-edge scattering at the defect induce $-y$ directional propagating modes on the right edge.
As the size of the defect increases as shown in Figs. \ref{propa}(d) and (e), it can be seen that the intensity of 
the skin mode on the right edge due to inter-edge scattering increases. 

the calculated transmittances ($|S_{21}|^2$ and $|S_{31}|^2$) and reflectances ($|S_{11}|^2$ and $|S_{41}|^2$) of the skin mode are shown 
in Fig. \ref{propa}(f). First, $|S_{31}|^2$ was always zero regardless of the defect 
size $d$, indicates that no skin mode propagating in the $+y$ direction appears on the right edge. 
Next, the change of $|S_{21}|^2$ is small when $d$ is small, and it begins 
to decrease when $d$ is about 7\% of the 
wavelength of the skin mode. In comparison, the change in $|S_{41}|^2$ was sensitive 
to the change in $d$, and the change in $|S_{41}|^2$ increased significantly when the 
defect size $d$ was increased. This is thought to be because the right-localized 
skin mode is amplified by inter-edge scattering due to the right diagonal gain in the $\mathrm{MT}$ symmetric
medium. Also, in Figs. \ref{propa}(c)-(e), no reflections of skin modes due to defect 
were observed on the left edge, and in Fig. \ref{propa}(f), $|S_{11}|^2$ is indeed zero in 
$d=0.005\ \mu\mathrm{m}, 0.01\ \mu\mathrm{m}$, and $0.02\ \mu\mathrm{m}$. However, as $d$ increases, $|S_{11}|^2$ also increases, 
which is thought to be due to the fact that some of the scattered waves are counted 
at Port1 as the right localized skin mode is amplified due to inter-edge scattering. 
When we performed the same calculations in the 50-100 THz frequency range, 
We have confirmed similar characteristics of transmittance and reflectance: $|S_{31}|^2$ is always zero, $|S_{21}|^2$ begins to decrease when $d$ is around 6-8\%, 
$|S_{41}|^2$ increases significantly with $d$, and $|S_{11}|^2$ is zero when $d$ is small and increases 
as $|S_{41}|^2$ increases.

Summarizing this section, we have described the results of our analysis of the unidirectional 
propagation of the skin mode and the properties of its scattering. 
In particular, we have shown that the scattering 
of the skin mode is peculiar, and that backscattering does not occur, but instead 
couples to the opposite edge.
Furthermore, the overall skin mode progression was rotation-like. 
This result could be used as a new method for generating optical orbital angular momentum (OAM).

\section{Circulating skin modes in $\mathcal{MT}$ symmetric case}\label{MT-circulating}
\subsection{Numerical results for closed square structures}\label{sec-square}

\begin{figure}[t]
  \includegraphics[width=0.8\columnwidth]{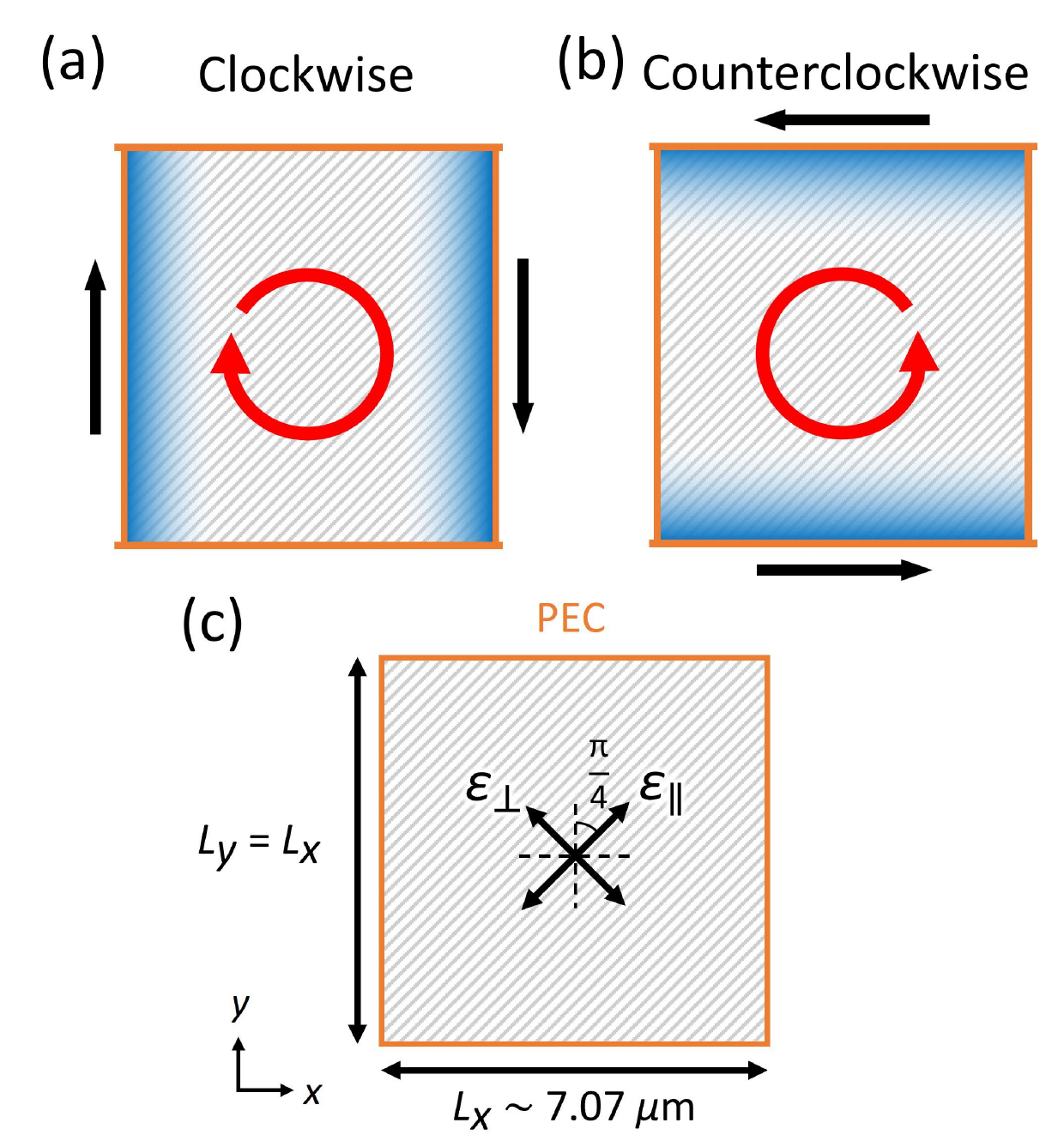}
  \caption{(a) and (b) show the modes that are expected to appear in the structure surrounded
   by the reflective boundaries. (a) Clockwise skin mode, localized on the left and right 
   edges. (b) Counterclockwise skin mode, localized at the top and bottom edges. (c) The structure 
   of the $\mathcal{MT}$ symmetric medium surrounded by PECs on the top, bottom, left and right edges, 
   used in the calculation.}\label{housin}
\end{figure}

Inspired by the inter-edge scattering of skin modes described in the previous section,
we consider a system with reflective bondaries not only at the left and right edges, but also at the top and bottom edges, forming a completely closed structure.
In this system, clockwise modes localized on the 
left and right edges may appear as shown in Fig. \ref{housin}(a), which may generate OAM. 
However, by providing reflective boundaries on the 
top and bottom as well, counterclockwise modes are also possible as shown in Fig. \ref{housin}(b). 
This is the first example of examining the use of NHSE to generate OAM.
First, we performed eigenmode calculations in a square structure with PECs on the top, bottom, left, and right 
edges of the $\mathcal{MT}$ symmetric medium same as in the previous section, as shown in Fig. \ref{housin}(c). 
Here, the $x$-directional width of the structure, $L_x$, is the same value as before, and $L_y (=L_x)$
represents the $y$-directional width of the structure. For the evaluation of the circulating modes, 
we also calculate the $z$ component of the orbital angular momentum per photon \cite{OAM}:
\begin{equation}
  \mathrm{OAM}=\frac{\iint \boldsymbol{E}^* \cdot\frac{\partial}{\partial \phi}\boldsymbol{E}\ dx dy}{\iint\boldsymbol{E}^*\cdot\boldsymbol{E}\ dx dy},
\end{equation}
where $\boldsymbol{E}$ is the transverse electric field and $\phi$ is the azimuth angle.

\begin{figure*}[t]
  \includegraphics[width=2\columnwidth]{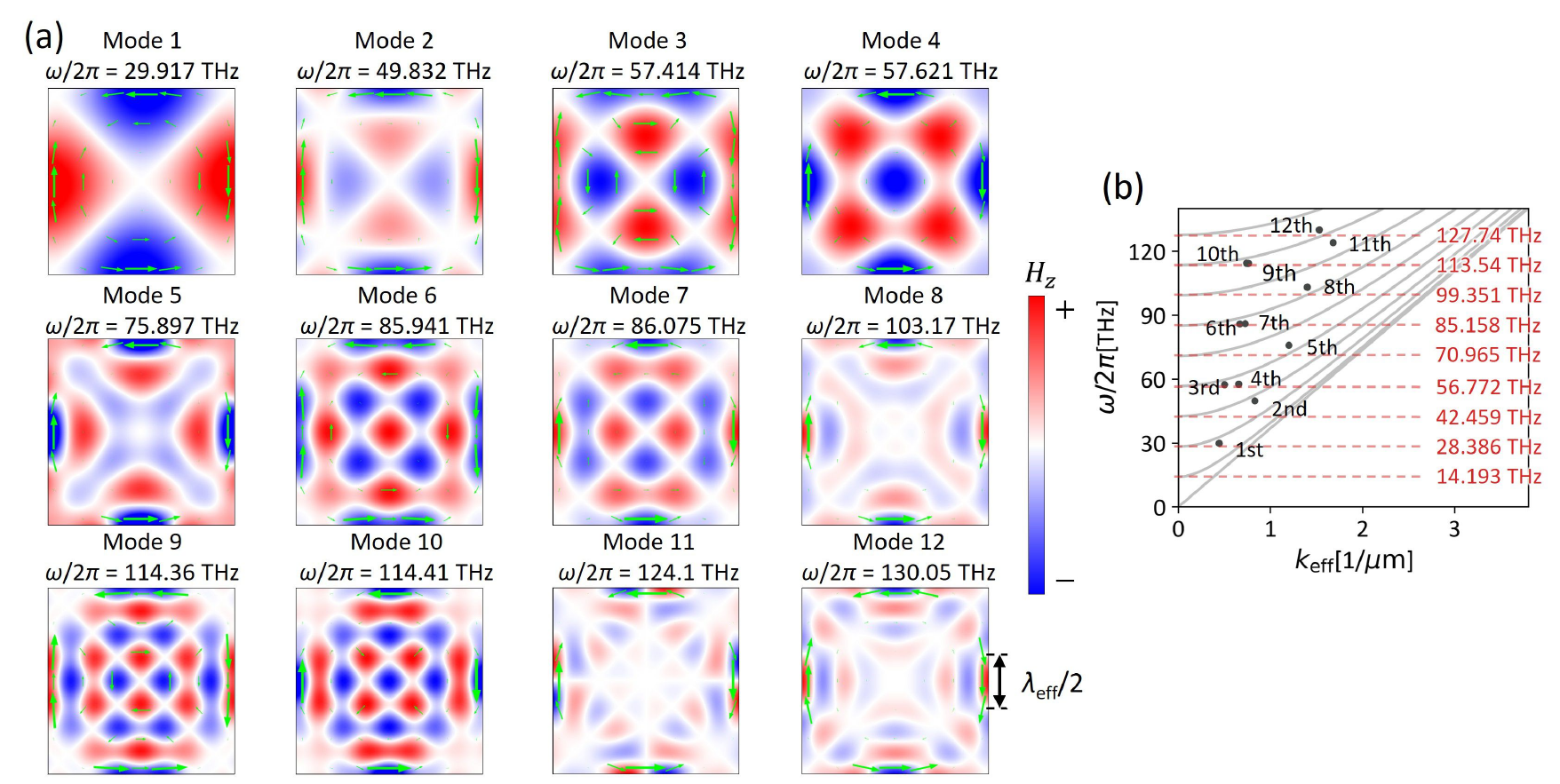}
  \caption{(a) $H_z$ distribution of eigenmodes from 1st to 12th order in square structures. 
  (b) The relationship between $k_\mathrm{eff}$ and $\omega/2\pi$ 
  for the modes of order 1 to 12. The gray lines show the relationship between 
  $k_y$ and $\omega/2\pi$ in the stripe system.
  The red dashed lines show the frequencies at which $k_y=0$.}\label{sq}
\end{figure*}
The $H_z$ distribution of eigenmodes are shown in Fig. \ref{sq}(a) up to the 12th order mode. 
The skin modes appear simultaneously at the top, bottom, left and right edges, 
which is particularly noticeable for the 2nd, 5th, 8th, 11th and 12th modes. 
The direction of propagation is $+y$ for the left edge and $-y$ for the right edge, 
and the overall direction is clockwise. Similarly, the skin mode propagates in 
the $-x$ direction at the top edge and in the $+x$ direction at the bottom edge, 
and the overall direction is counterclockwise. In the square structure, these 
rotations are balanced and the net OAM is zero.

Here, we estimate the effective wavelength $\lambda_\mathrm{eff}$, where we define $\lambda_\mathrm{eff}$ 
as twice the distance between nodes in the propagation direction of the skin mode, as shown in the 12 th mode in Fig. \ref{propa}(a). 
The effective wavenumber is obtained as $k_\mathrm{eff}=2\pi/\lambda_\mathrm{eff}$, and plotted 
it over the $k_y$-$\omega/2\pi$ relation in the stripe system (Fig. \ref{MTskin}(d)) in Fig. \ref{sq}(b).
It can be seen that the relationship between $k_\mathrm{eff}$ and $\omega/2\pi$ is almost distributed 
along the frequency dispersion of the stripe system.
Therefore, skin modes corresponding to the eigenmodes in the stripe system are 
expected to appear in the finite system as well. Indeed, corresponding to the fact 
that the localization of eigenmodes of the stripe system becomes stronger as $k_y$ 
is larger due to the term $e^{iqk_y x}$ in Eqs. (\ref{Ey})-(\ref{Ex}),
the localization is stronger in the 2nd, 5th, 8th, 11th, and 12th 
modes where $k_\mathrm{eff}$ is large. Here, since $q=0.5774$, the localization lengths of 
the 2nd, 5th, 8th, 11th, and 12th modes are obtained from each $k_\mathrm{eff}$ 
as $|\mathrm{Im}(qk_\mathrm{eff})|= 2.09\ \mu\mathrm{m}$, $1.44\ \mu\mathrm{m}$, $1.24\ \mu\mathrm{m}$, $1.03\ \mu\mathrm{m}$, and $1.13\ \mu\mathrm{m}$, respectively. 
These are roughly consistent with the distribution of $H_z$ obtained from the simulation.

Furthermore, from Eq. (\ref{f-k}), the frequency at which $k_y=0$, i.e., 
the eigenmode is delocalized in the stripe system, is obtained as 
$(c\sqrt{\eta_{yy}} /2L_x )n=14.193n$ THz.
Figure \ref{sq}(a) shows that the 
localization is weaker in the eigenmodes at frequencies near $14.193n$ THz. 
This is also a confirmation that the modes corresponding to the stripe 
system appear in the finite system.
\begin{figure}[t]
  \includegraphics[width=\columnwidth]{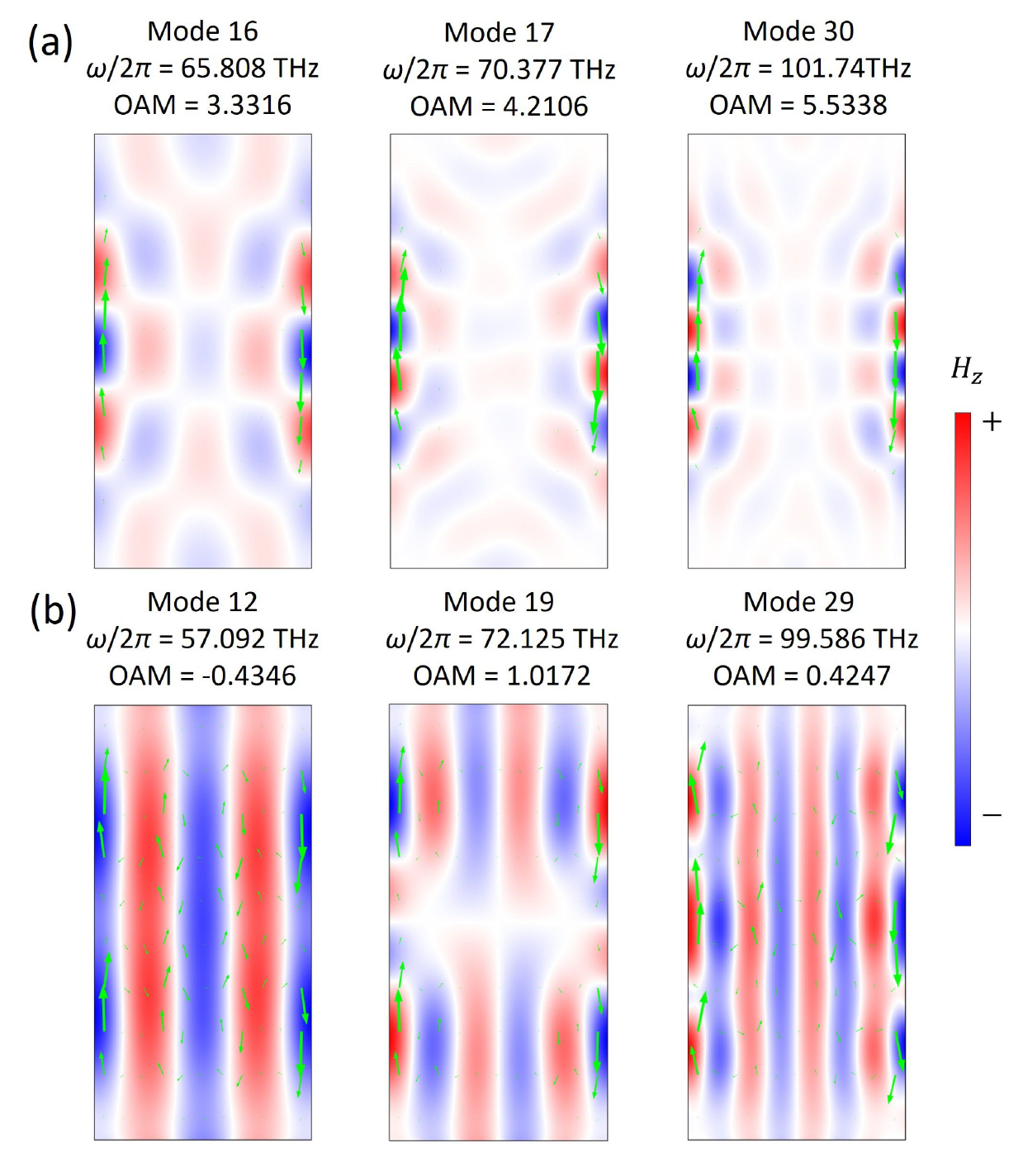}
  \caption{$H_z$ distribution of eigenmodes in a rectangular structure of $L_y=2L_x$.
  (b) shows 16th-, 17st-, and 30st- modes as examples of strongly localized modes.
  (a) shows 12th-, 19th-, and 29th- modes as examples of weakly localized modes.}\label{rec}
\end{figure}
\begin{figure}[t]
  \includegraphics[width=1\columnwidth]{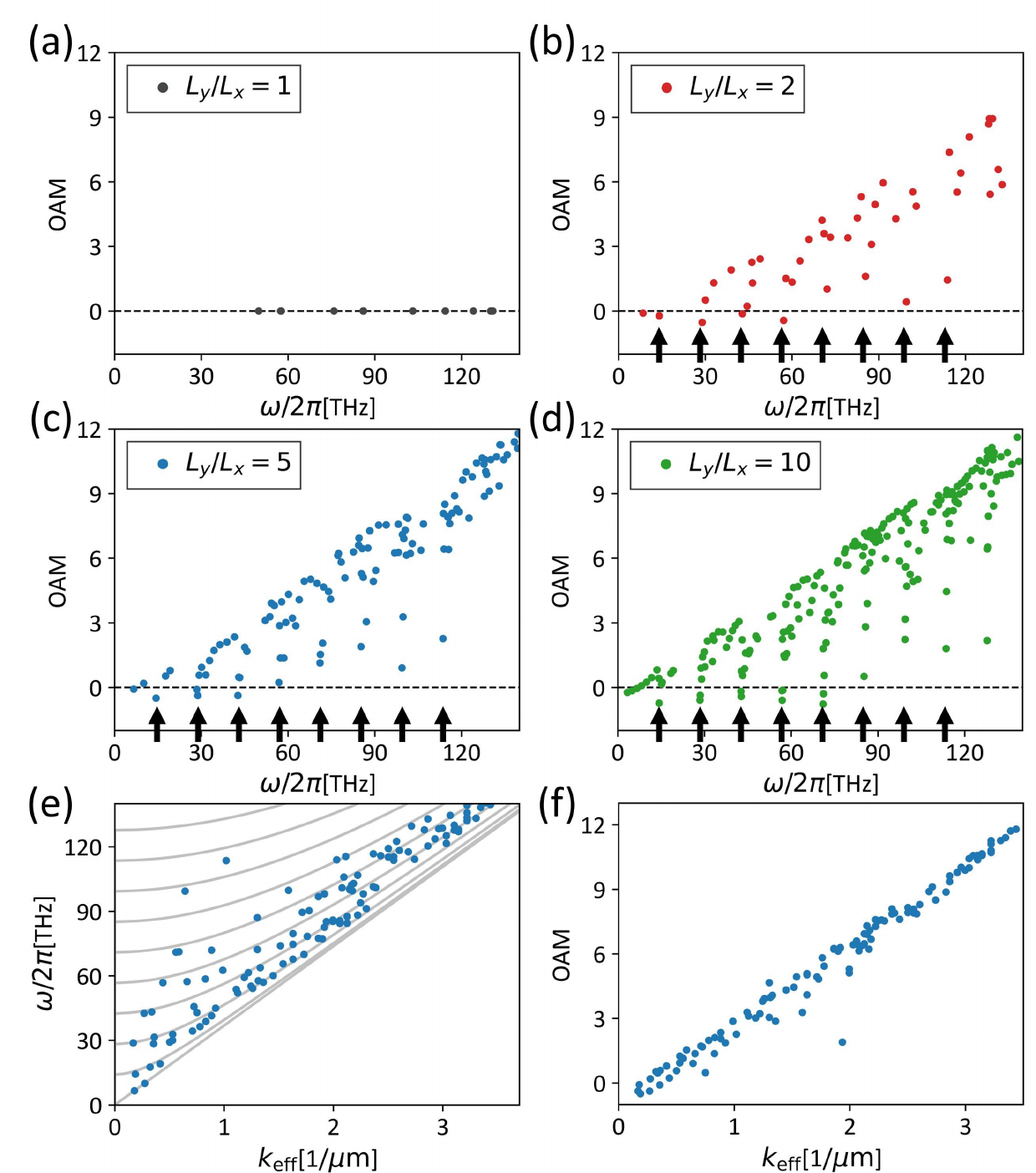}
  \caption{(a)(b)(c)(d) Relationship between calculated eigenfrequencies $\omega/2\pi$ and OAM for $L_y/L_x=1, 2, 5$, and $10$, respectively.
  In (b)-(d), the frequencies at which the OAM is close to zero are indicated by black arrows.
  (e) Relationship between the effective wavenumber $k_\mathrm{eff}$ and $\omega/2\pi$. 
  The gray lines show the relationship between $k_y$ and $\omega/2\pi$ in the stripe system.
  (f) Relationship between $k_\mathrm{eff}$ and OAM.}\label{f-oam}
\end{figure}

\subsection{Numerical results for closed rectangular structures}\label{sec-rec}
Next, we performed the eigenmode calculations with a rectangular structure 
with $L_y=2L_x$ in order to break the balance between clockwise 
and counterclockwise rotation. Some of the modes with particularly strong 
localization are shown in Fig. \ref{rec}(a). In this case, only clockwise modes 
localized at the left and right boundaries strongly appear, 
successfully breaking the rotation balance. This also resulted in a non-zero 
finite value of OAM. As in the square case, modes with weak localization 
appeared at frequencies around $14.193n$ THz, resulting in small values 
of OAM as shown in Fig. \ref{rec}(b).
In some modes, such as mode 12 in Fig. \ref{rec}(b), the OAM was negative, 
which will be discussed later.

Now consider the case where $L_x \sim 7.07 \mu\mathrm{m}$ is fixed and $L_y$ is varied. 
The results of plotting the relationship between eigenfrequencies and OAM for each $L_y/L_x$ 
are shown in Figs. \ref{f-oam}(a)-(d). For a square ($L_y/L_x=1$), the all OAMs are zero, but as $L_y/L_x$
increases, the OAMs increase, and it is found that the OAMs saturate around $L_y/L_x=5$.
The increase in OAM with $L_y/L_x$ is caused by the fact that in closed structures skin modes corresponding to eigenmodes in the stripe system appear. 
As in the case of the square structure, we estimated the effective wavelength 
$\lambda_\mathrm{eff}$ and the effective wavenumber $k_\mathrm{eff}$ for $L_y/L_x=5$, 
and plotted the $k_\mathrm{eff}-\omega/2\pi$ relationship 
in Fig. \ref{f-oam}(e). 
Each point is almost distributed along the $k_y$-$\omega/2\pi$ curve in the stripe system, 
which is especially noticeable in the region where $k_\mathrm{eff}$ is small, indicating 
that the skin mode in the closed system corresponds 
to the skin mode in the stripe system. In addition, the distribution of $\omega/2\pi$-OAM in 
the rectangular structure has a comb-like shape with OAMs periodically close to zero (Shown by the black arrows in Fig. \ref{f-oam}), 
which is similar to the $\omega/2\pi$-$k_y$ relationship in the stripe system. 
Therefore, when we plot the $k_\mathrm{eff}$-OAM relationship, especially for the case 
$L_y/L_x=5$, we obtain a roughly linear relationship as shown in Fig. \ref{f-oam}(f).

Here, skin modes at the left and right edges are formed by repeated reflections at 
the left and right edges while the light undergoes anisotropic loss or gain. Thus, 
in a structure with small $L_y/L_x$, when $k_\mathrm{eff}$ is large, skin modes can not be formed at 
the left and right edges, limiting the value of $k_\mathrm{eff}$ that can be taken. When $L_y/L_x$ 
is large, skin modes can be formed on the left and right edges even when $k_\mathrm{eff}$ is large. 
Therefore, as the aspect ratio of the structure is increased, the OAM proportional 
to $k_\mathrm{eff}$ is also expected to increase.

Moreover, when the aspect ratio becomes high enough, skin modes corresponding 
to eigenmodes at any points on the frequency dispersion of the stripe system 
(Fig. \ref{MTskin}(d)) can be formed.
As mentioned above, the skin mode in the closed system correspond to the skin mode in the stripe system. Therefore, no matter how much 
the aspect ratio of the rectangle is increased, modes with wavenumbers in the area 
below the $n=0$ line in Fig. \ref{MTskin}(d) will not appear, $k_\mathrm{eff}$ 
will saturate, and the OAM proportional to $k_\mathrm{eff}$ will also saturate.

Now we discuss the fact that the OAM is negative in some modes, as can be seen in Figs. \ref{f-oam}(b)-(d).
As mentioned above, OAM and $k_\mathrm{eff}$ are proportional, and when OAM is small, 
$k_\mathrm{eff}$ is also small. This results in a smaller contribution from the clockwise 
wave traveling along the edge. Instead, the contribution of counterclockwise 
traveling waves inside the medium becomes relatively larger (see videos in the supplemental material). 
Therefore, the value of OAM becomes negative in some case close to the zero-$k_\mathrm{eff}$ condition.
\begin{figure*}[ht]
  \includegraphics[width=2\columnwidth]{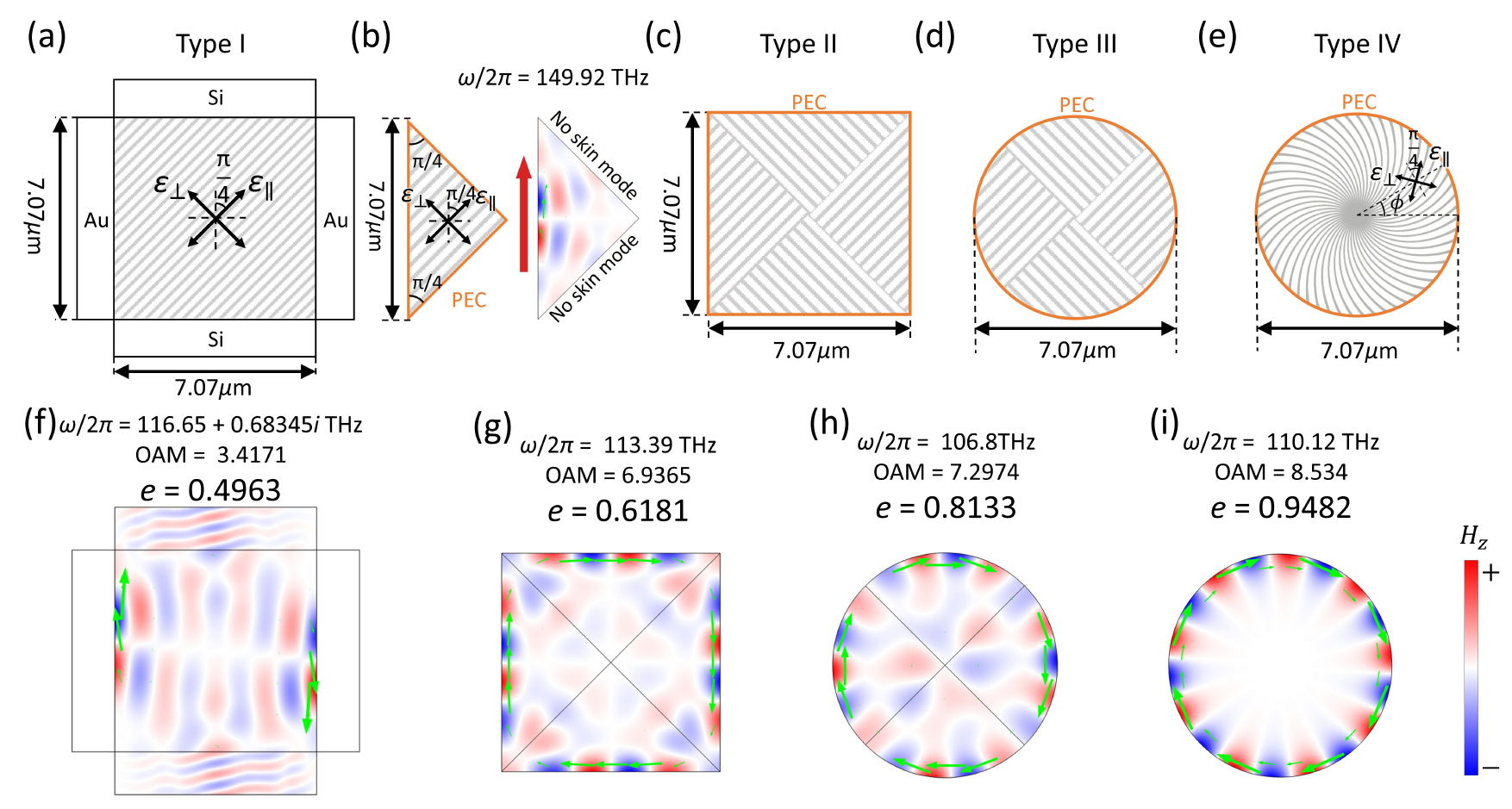}
  \caption{(a) Square $\mathcal{MT}$ symmetric medium sandwiched between Au layers 
  in the x-direction and Si layers in the y-direction.
  (b) Isosceles right triangular $\mathcal{MT}$ symmetric medium and the eigenmodes in this structure.
  (c) Structure consisting of four triangular $\mathcal{MT}$ symmetric media in (b). 
  (d) Circular structure consisting of four fan-shaped $\mathcal{MT}$ symmetric mediums.
  (e) Circular structure consisting of four fan-shaped $\mathcal{MT}$ symmetric media. 
  The dielectric constants are $\varepsilon_{xx}=\varepsilon_{yy}=\varepsilon_{\perp} \cos^2 (\phi+\pi/4)
  +\varepsilon_{\parallel}\sin^2(\phi+\pi/4)$,
  $\varepsilon_{xy}=\varepsilon_{yx}= (\varepsilon_{\perp}-\varepsilon_{\parallel})
  \sin(\phi+\pi/4)cos(\phi+\pi/4)$, where $\varepsilon_\parallel=1.673+0.96593i, \varepsilon_\perp=1.673-0.96593i$.
  (f)(g)(h)(i) Eigenmodes in the structures (a), (c)-(e), respectively.}\label{kouritsu}
\end{figure*}
\subsection{Generation of high-efficiency circulating mode}\label{sec-kouritsu}
Next, we explore the generation of more efficient circulating modes. 
To evaluate the efficiency of the OAM, we define the efficiency $e$
as the normalization of the OAM with the number of effective wavelengths of the circulating 
mode contained in the perimeter of the structure:
\begin{equation}
  e = \frac{\mathrm{OAM}}{L/\lambda_\mathrm{eff}}
\end{equation}

where $L$ is the perimeter of the structure. Here, for an ideal ciruculating mode such as the Laguerre-Gaussian mode \cite{LG}, the OAM efficiency is $e=1$.
As an example, the efficiency at mode 30 of rectangular structure with $L_y/L_x=2$ 
(Fig. \ref{rec}(a)) is calculated to be $e=0.4246$. This is a mode with higher efficiency in the rectangular 
structure, but still low.
Here, we seek a structure that can 
achieve higher efficiencies.

First, instead of breaking the equivalence of edge lengths as in the 
rectangular structure, we brake the equivalence of interface types 
as shown in Fig. \ref{kouritsu}(a) and performed eigenmode calculations in a square 
$\mathcal{MT}$ symmetric medium sandwiched between Au layers in the $x$ direction and Si layers in 
the $y$ direction. We refer to this structure as Type I structure. 
The calculated $H_z$ distribution is shown in Fig. \ref{kouritsu}(f). 
In this case, as in the rectangular structure, the left-right localized 
clockwise modes appear strongly and OAM is generated. Here, the efficiency
of the OAM is $e=0.4963$, which is higher than that of the rectangular structure.

Next, as shown in Fig. \ref{kouritsu}(b), we performed calculations with a isosceles right 
triangle structure and found that skin modes do not appear on the diagonal edges, 
and skin modes propagating in the $+y$ direction appear only on the perpendicular edges. 
Here, this 45-degree diagonal edge is parallel to the mirror symmetry plane of the dielectric tensor. 
The fact that skin modes do not appear on the edge parallel to the mirror symmetry plane of the system is known as a 
property of the geometry dependent skin effect (GDSE) \cite{uni}. This is a type of NHSE in fully closed 
boundary systems, and GDSE is thought to be appearing in the medium used in this study.
Therefore, in order to generate a one-directional rotation on the edge, we considered 
a square structure consisting of four triangular $\mathcal{MT}$ symmetric media, as shown in 
Fig. \ref{kouritsu}(c). We reffer to this structure as Type II structure. 
The calculated $H_z$ distribution for this 
structure is shown in Fig. \ref{kouritsu}(g). We succeeded in generating skin modes 
that circulate not only on the left and right edges but also around the perimeter 
in one direction. Even when the structures are combined, the properties of GDSE are considered valid, and
we expect that the one-directional circulation localized on the perimeter of such a
structure is obtained.
Also, in this case, the efficiency of the OAM is $e=0.6181$, which 
is higher than that of Type I structure.
In addition, because of the nature of GDSE, the conditions for skin modes to appear is satisfied
except at 45-degree diagonal edges. Thus, we also considered the structure shown in the Fig. \ref{kouritsu}(d), 
which consists of four fan-shaped $\mathcal{MT}$ symmetric media. We refer to this structure as Type III structure.
In this structure, $e=0.8133$ is obtained, which is a higher efficiency than that of the circulating mode in Type II structure.

In addition, we calculated in the circular structure 
with anisotropic dielectric permittivity with $\mathcal{MT}$ symmetric anisotropy continuously 
rotated around the center of the circle, as shown in Fig. \ref{kouritsu}(e). Each 
component of the dielectric constant tensor is  $\varepsilon_{xx}=\varepsilon_{yy}=\varepsilon_{\perp} \cos^2 (\phi+\pi/4)
+\varepsilon_{\parallel}\sin^2(\phi+\pi/4)$,
$\varepsilon_{xy}=\varepsilon_{yx}= (\varepsilon_{\perp}-\varepsilon_{\parallel})
\sin(\phi+\pi/4)cos(\phi+\pi/4)$. We refer to this structure as Type IV strucuture.
In the case of the Type III structure, the appearance of skin modes is weak near the point 
where the tangent line of the circle is 45 degrees diagonal, due to the characteristics of GDSE. 
On the other hand, in the Type IV structure, skin modes can exist for all angles on the circle, 
which is expected to result in better efficiency $e$.
The calculated $H_z$ distribution is shown in Fig. \ref{kouritsu}(i).
In this case, as expected, a skin mode circulating in one direction appeared on the perimeter.
Also, the OAM efficiency is $e=0.9482$, 
which is the highest efficiency of the circulating modes in the structures considered in this section.

\subsection{Comparison of purity of circulating modes}
Next, we discuss the purity of the circulating mode using the following equation \cite{spec}. Hereafter we will refer to this as the OAM spectrum.
\begin{equation}
  F(l)=\frac{|d_l|^2}{\sum_{l'}|d_{l'}|^2},
\end{equation}
\begin{equation}
  d_l=\int_0^{2\pi}\int_0^\infty H_z e^{-il\phi}rdr d\phi,
\end{equation}
\begin{figure}[t]
  \includegraphics[width=1\columnwidth]{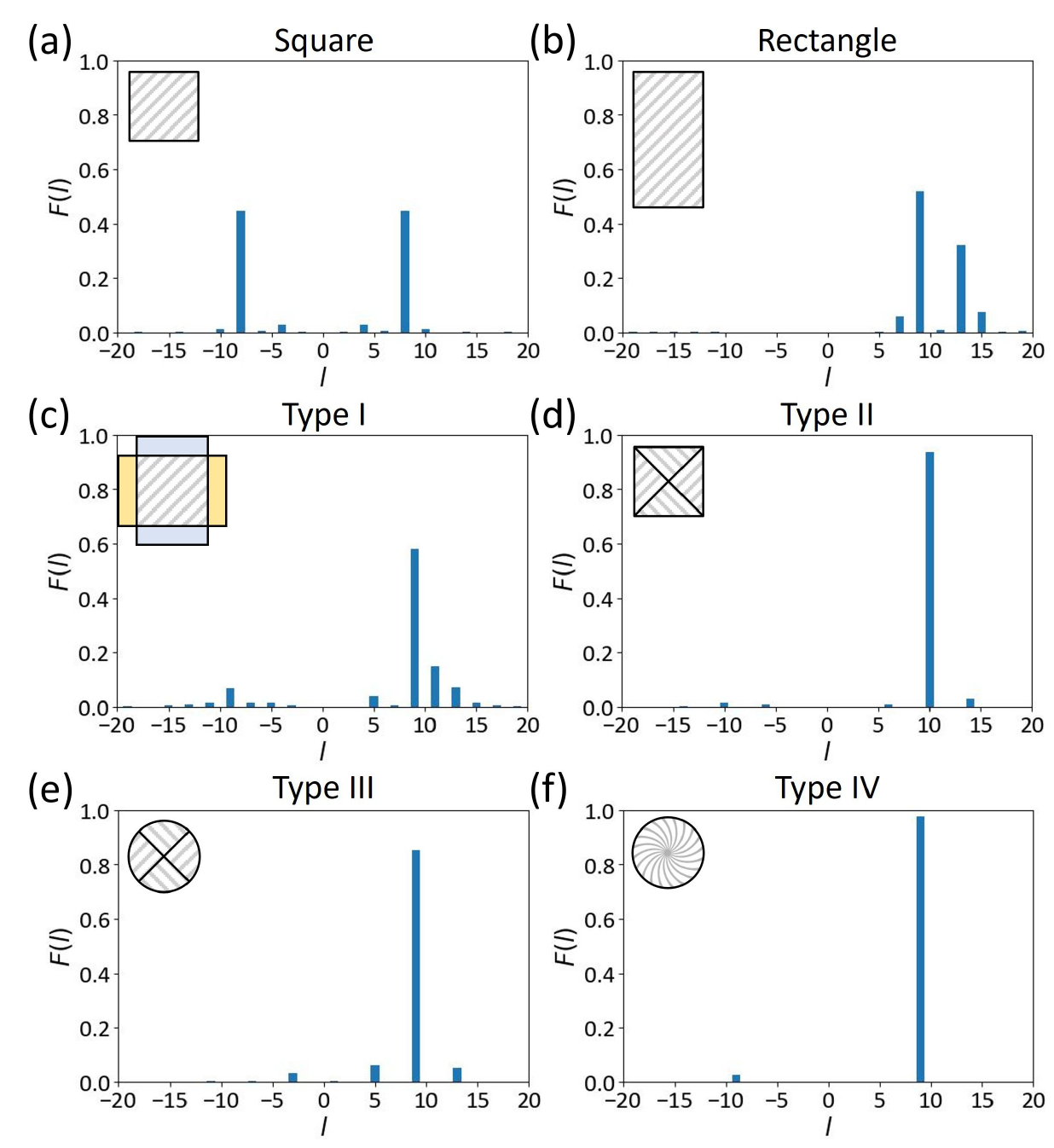}
  \caption{(a)(b)(c)(d)(e)(f) OAM spectrum of mode 8 in Fig. \ref{sq}(a), 
  mode 30 in Fig. \ref{rec}(a), and modes shown in Fig. \ref{kouritsu}(f), (g), (h), 
  and (i), respectively. A schematic of the structure is shown in the upper 
  left corner of each graph.}\label{spe}
\end{figure}
where $l$ is the azimuthal mode number. $l$ can be any integer, with positive $l$ 
corresponding to a clockwise mode and negative $l$ to a counterclockwise mode.
Here the OAM spectrum is the content of each basis when $H_z$ 
is expanded into a basis of $e^{il\phi}$. Note that for a pure circulating mode with a spatial distribution of the field 
that depends on $e^{il\phi}$ ($l\neq0$) such as Laguerre-Gaussian mode, the OAM spectrum is $F(l) = 1$.

Figure \ref{spe} shows the results of the OAM spectrum calculations. 
First, for the mode 8 in the square structure (Fig. \ref{sq}(a)), the OAM spectrum 
has a symmetric distribution (Fig. \ref{spe}(a)). This is corresponding to the 
fact that the clockwise and counterclockwise modes are in balance and the 
OAM is zero. Next for the mode 30 in the rectangular structure (Fig. \ref{rec}(a)) and the mode in Type I structure (Fig. \ref{kouritsu}(a)),
the left-right symmetry of the OAM spectrum is broken and the OAM spectrum is strongly 
distributed on the positive side of $l$ (Figs. \ref{spe}(b) and (c)). This is corresponding to the strong appearance of 
clockwise modes localized on the left and right sides. 
Furthermore, for the modes in Type II structure (Fig. \ref{kouritsu}(g)) 
and Type III structure (Fig. \ref{kouritsu}(h)), 
the variation in the distribution of the OAM spectrum is smaller, 
corresponding to the high OAM efficiency obtained in the previous section. 
In particular, it is more pronounced for Type III structure. The mode in Type IV structure (Fig. \ref{kouritsu}(i)) shows the smallest 
variation in the distribution of the OAM spectrum and the highest purity. 
This is also consistent with the highest efficiency obtained in the previous section 
for this structure.

\begin{figure*}[ht]
  \includegraphics[width=2\columnwidth]{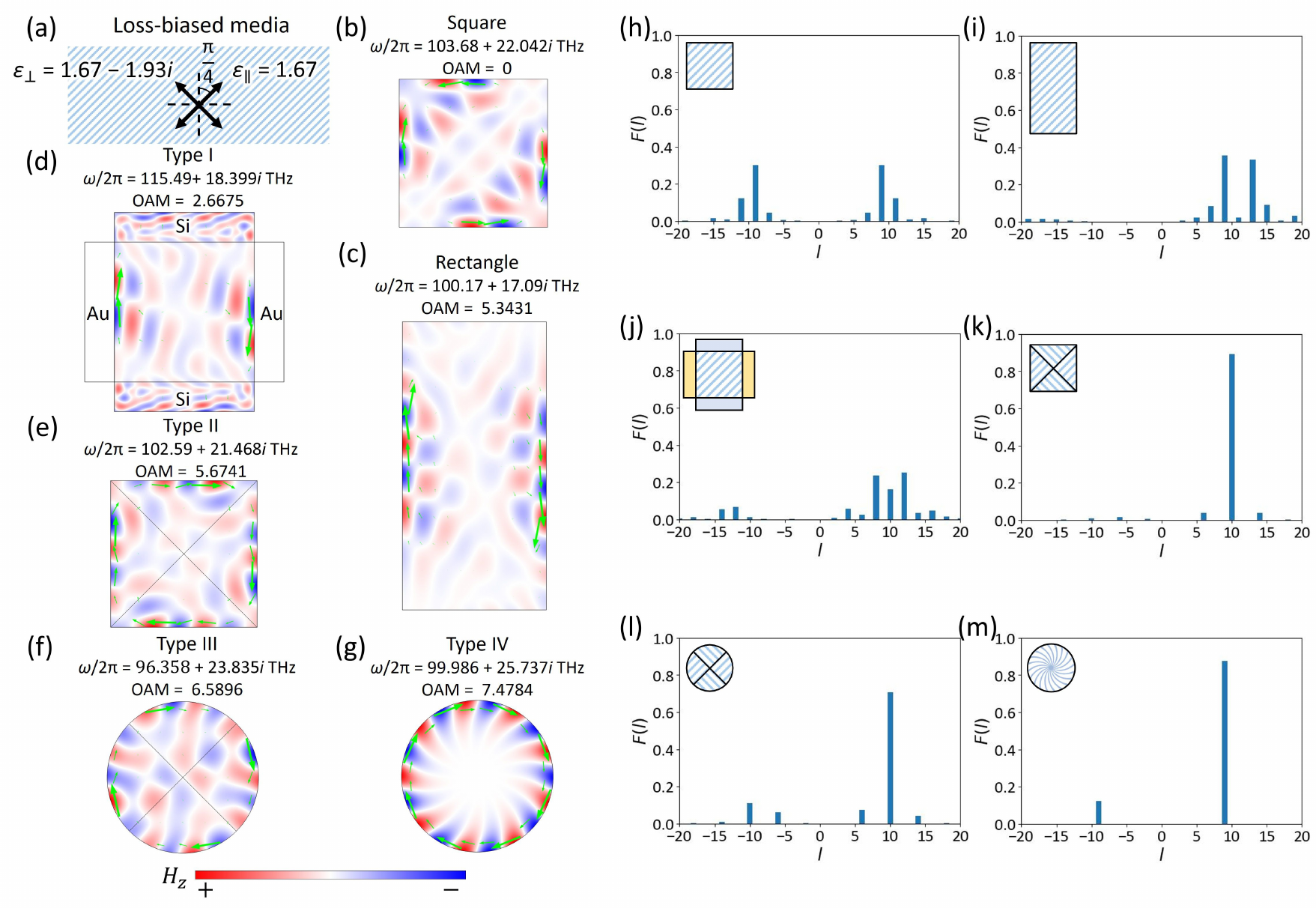}
  \caption{(a) Schematic of loss- biased media.
  (b)(c)(d)(e)(f)(g) The eigen modes in the square structure, the rectangular 
  structure ($L_y/L_x=2$), the structure sandwiched between Au and Si layers, 
  the structure combining right triangle, the structure combining fan shape, 
  and the structure with continuously rotated anisotropy around the center, respectively.
  (h)(i)(j)(k)(l)(m) Fourier spectra of modes shown in (b)-(g), respectively.}\label{lossbias}
\end{figure*}

\begin{table*}[ht]
  \caption{Comparison of OAM efficiency in the loss-gain balanced case and in the Loss bias case}\label{hyou1}
  \includegraphics[width=1.5\columnwidth]{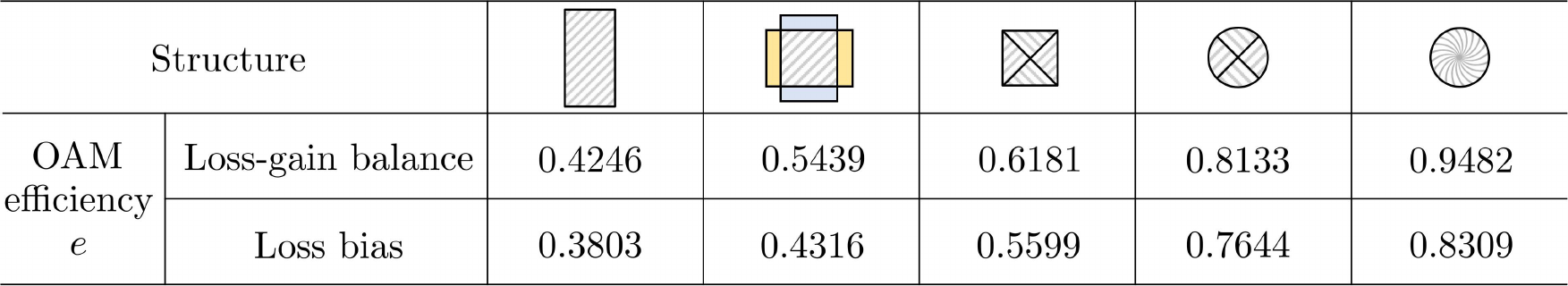}
\end{table*}
\section{Circulating skin modes in loss-biased case}\label{loss-circulating}

So far, we have considered circulating modes in a medium in which gain and loss 
are balanced. However, it is difficult to realize a medium with large gain and 
balanced gain and loss in a real material. In this section, we consider circulating 
modes in loss-biased media with a view to demonstrating them experimentally. 
Here, as in Fig. \ref{lossbias}(a), we use an anisotropic loss-biased medium with 
dielectric constants $\varepsilon_{xx}=\varepsilon_{yy}=1.673-0.96593i$ and
$\varepsilon_{xy}=\varepsilon_{yx}=-0.96593i$, given by Eq. (\ref{eq1}), 
$\varepsilon_{\parallel}=1.673$, and $\varepsilon_{\perp}=1.673-1.93186i$. 
These are obtained by increasing the loss while holding the contrast of the imaginary part 
of the $\mathcal{MT}$ symmetric dielectric constants ( $\varepsilon_{\parallel}=1.673+0.96593i$ and $\varepsilon_{\perp}=1.673-0.96593i$) that we have used.
Then, we will replace the structures in the previous section 
with this medium.

Figures \ref{lossbias}(b)-(d) show the calculated $H_z$ distributions for each structure. 
Note that these are the modes whose effective wavenumbers are close to those of the mode 8 in Fig. \ref{sq}, the mode 30 in Fig. \ref{rec}, 
and the modes in Figs. \ref{kouritsu}(f)-(i), respectively.
First, in the square structure, clockwise modes localized on the 
left and right edges and counterclockwise modes localized on the 
top and bottom edges appeared simultaneously, and they were in balance, 
resulting in zero OAM (Fig. \ref{lossbias}(b)). Corresponding to this, the OAM spectrum was symmetrically distributed(Fig. \ref{lossbias}(h)). 
Next, in the rectangular structure ($L_y/L_x=2$) and Type I structure, clockwise modes strongly appeared (Figs. \ref{lossbias}(c) and (d)), 
and the OAM spectrum was asymmetric (Figs. \ref{lossbias}(i) and (j)).

In addition, in Type II-IV structures, circulating mode traveling in one direction around the perimeter 
of the structure appear (Figs. \ref{lossbias}(e)-(g)), and the variation of the OAM spectrum is smaller 
than that of the rectangular and Type I structures (Figs. \ref{lossbias}(k)-(m)). 
In particular, Type IV structure has the purest circulating mode with the smallest variation of the OAM spectrum of the modes in the
square, rectangle, and type I-IV structures. 
These are the same trends as when using $\mathcal{MT}$ symmetric media.

Also, the calculated efficiencies of the OAM for each structure are shown in Table \ref{hyou1}. 
Note that the efficiencies in the $\mathcal{MT}$ symmetric case are also shown for comparison.
This shows that the efficiency increases in the order of the rectangular, Type I, II, III and IV structures, 
which is also the same trend as in the $\mathcal{MT}$ symmetric case.

Then, we compare the OAM efficiency in the $\mathcal{MT}$ symmetric case and the loss-biased case.
Table \ref{hyou1} shows that for each structure, the OAM efficiency in the loss-biased case is lower 
than in the $\mathcal{MT}$ symmetric case. This reduction is slight for the loss introduced, 
suggesting that it is easy to realize. Here, possible reasons for this decrease in efficiency are, first, 
the imbalance between loss and gain, which tilts the propagation direction of the circulating mode 
toward the azimuthal direction, and second, the longer localization length of the skin mode than in the $\mathcal{MT}$ symmetric case, which increases 
its contribution to the counterclockwise circulation, resulting in a decrease in net OAM. We discuss this in detail in Appendix B and C.

As described above, OAM is feasible even with a loss-biased medium.
The OAM and OAM spectral features for each structure were similar to those obtained with $\mathcal{MT}$ symmetric media, and the OAM decrease was slight for the introduced loss.
However, the imaginary part of the eigenfrequencies increase.

\section{Summary and discussion}
We have investigated numerically for the first time the propagation 
of NHSE in two-dimensional reciprocal uniform 
media. In the simulation of the propagation of skin modes with the 
$\mathcal{MT}$ symmetric medium, we found an interesting phenomenon: the skin 
modes jump between opposite edges due to scattering and travel in 
opposite directions to each other.

Applying this inter-edge coupling, we have investigated a square structure surrounded by PECs.
We found that clockwise modes localized on 
the left and right edges and counterclockwise modes localized on the top 
and bottom edges appeared simultaneously, and the OAM became zero by 
balancing them. However, in a rectangular structure or a structure sandwiched 
between different media, such as Au layers on the left and right and Si 
layers on the top and bottom, one rotation was selectively exhibited and 
a non-zero OAM was obtained. In other words, finite OAM can be generated by breaking the balance between the horizontal and vertical boundaries.
In addition, we confirmed that high efficiency 
and high purity circulating modes can be achieved by a structure 
combining right-angle triangles and fan- shapes. We examined these structures inspired by the 
fact that skin modes do not appear at the 45-degree diagonal edges in the 
right triangle structure. In Ref. \cite{uni}, the geometry dependence of the 
skin state in two-dimensional lattice models is discussed. 
As a property of geometry-dependent NHSE (GDSE), 
the skin state appearing at each edge in a fully closed system corresponds 
to the skin state in an infinitely long stripe geometry parallel to that edge and
that the skin state does not appear on edges parallel to the mirror symmetry plane of the system.
The former property was discussed in section \ref{sec-square} and \ref{sec-rec} and the latter in section \ref{sec-kouritsu}. 
Thus, it can be assumed that GDSE is also appearing in the system we have considered.
Therefore, it can be said that our results confirm for the first time the application of GDSE as the OAM generation.
Furthermore, we considered a medium with continuously rotating anisotropy 
and obtained the most efficient and purest circulating modes.

We also replaced the medium with an anisotropic loss-bias medium and performed 
the same calculations and found that the circulating mode can still be generated 
in this case. This result suggests experimental demonstrability, since unlike 
the $\mathcal{MT}$ symmetric medium, no gain is required to construct this medium and it is easy to realize.
Our work also paves the way for the application of NHSE as micro-sized optical devices manipulating OAM, 
such as OAM light emitters and OAM-dependent transmission or reflection devices.

\begin{acknowledgements}
  This work was supported by the Japan Society for
  the Promotion of Science (Grant number JP20H05641,
  JP21K14551, 24K01377, 24H02232 and 24H00400).
  
\end{acknowledgements}

\section*{Appendix A: Composition of $\mathcal{MT}$ symmetric dielectric constants with multilayers}
\renewcommand{\theequation}{A\arabic{equation}}
\renewcommand{\thefigure}{A\arabic{figure}}
\setcounter{equation}{0}
\setcounter{figure}{0}
In this section, we consider composing $\mathcal{MT}$ symmetric dielectric constants with multilayers.
We consider a multilayer structure tilted by 45 degrees as shown in Fig. \ref{a1}, 
where $a$ is the lattice constant, $\varepsilon_1$ is the dielectric constant of medium 1, 
$\varepsilon_2$ is the dielectric constant of medium 2, and $f$ is the ratio of the thickness 
of medium 1 to the lattice constant $a$. When $a$ is sufficiently shorter than the wavelength 
of light, by the effective medium approximation, the dielectric constants $\varepsilon_\parallel$ 
in the direction parallel to the multilayer and $\varepsilon_\perp$ in the direction perpendicular 
to it can each be expressed as follows:
\begin{equation}
  \varepsilon_\parallel=f\varepsilon_1+(1-f)\varepsilon_2,\ \varepsilon_\perp=\frac{\varepsilon_1\varepsilon_2}{(1-f)\varepsilon_1+f\varepsilon_2}.\label{MLapprox}
\end{equation}
\begin{figure}[t]
  \includegraphics[width=0.6\columnwidth]{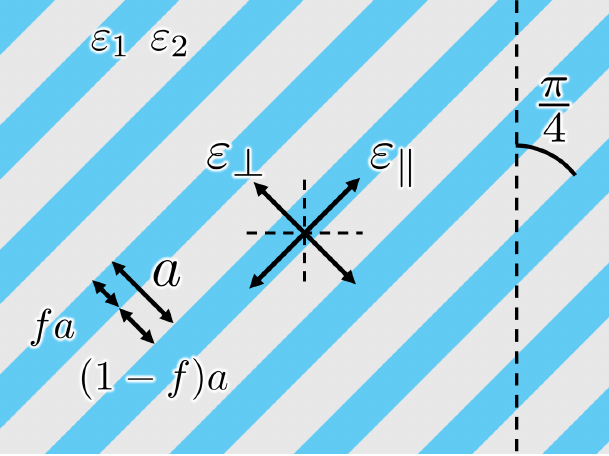}
  \caption{Schematic of multilayer structure}\label{a1}
\end{figure}
Here, the dielectric constants of mediums 1 and 2 are written as
\begin{equation}
  \varepsilon_1=A_1e^{i\phi_1},\ \varepsilon_2=A_2e^{i\phi_2}\ (-\pi<\phi_1,\phi_2<\pi).\label{eps12-def}
\end{equation}
Then, we find the conditions for $A_1, A_2, \phi_1$, and $\phi_2$ 
under which the $\mathcal{MT}$ symmetry holds. As mentioned in Section \ref{sec2-a}, 
$\mathcal{MT}$ symmetry is satisfied when $\varepsilon_\parallel=\varepsilon_\perp^*$. Substituting Eq. (\ref{eps12-def}) 
into this equation and multiplying both sides by $(1-f)\varepsilon_1+f\varepsilon_2$, 
we obtain the following equation:
\begin{equation}
  f(1-f)\qty(|\varepsilon_1|^2+|\varepsilon_2|^2)+f^2\varepsilon_1^*\varepsilon_2+(1-f)^2\varepsilon_1\varepsilon_2^*=\varepsilon_1^*\varepsilon_2^*.\label{f-eps12}
\end{equation}
Substituting Eq. (\ref{eps12-def}) and considering the imaginary part, we obtain
\begin{eqnarray}
  -f^2A_1A_2\sin(\phi_1-\phi_2)+&&(1-f)^2A_1A_2\sin(\phi_1-\phi_2)\nonumber\\
  &&=-A_1A_2\sin(\phi_1+\phi_2).
\end{eqnarray}
By rearranging this equation, the condition on $\phi_1,\phi_2$ can be obtained as
\begin{equation}
  \phi_2=\tan^{-1}\qty(\frac{f-1}{f}\tan\phi_1).\label{phi1-phi2}
\end{equation}
Since $(f-1)/f<0$, $\mathrm{sign}(\phi_1) \neq \mathrm{sign}(\phi_2)$. 
Thus, we can confirm that both gain and loss media are required to compose 
$\mathcal{MT}$ symmetric dielectric constants 
by the effective medium approximation of the multilayer. 
Next, considering the real part of Eq. (\ref{f-eps12}), we obtain
\begin{eqnarray}
  f(1-f)(A_1^2+A_2^2)&&+f^2A_1A_2\cos(\phi_1-\phi_2)\nonumber\\
  +(1-f)^2\cos(\phi_1-\phi_2)&&=A_1A_2\cos(\phi_1+\phi_2).
\end{eqnarray}
By rearranging this equation, the conditions on $A_1$ and $A_2$ are given by
\begin{equation}
  A_2=\frac{1}{2}\qty(X\pm\sqrt{X^2-4})A_1,
\end{equation}
\begin{equation}
  X=\frac{\cos(\phi_1+\phi_2)-\qty(f^2+(1-f)^2)\cos(\phi_1-\phi_2)}{f(1-f)}.
\end{equation}

In particular, considering the case $f=1/2$, from Eq. (\ref{phi1-phi2}), $\phi_2=-\phi_1$. 
Furthermore, by setting $\phi_2=-\phi_1=\pi/4, A_1=1$, and $A_2=2+\sqrt{3}$, 
the dielectric constants of each medium are obtained as $\varepsilon_1=0.70711-0.70711i$ and 
$\varepsilon_2=2.639+2.639i$. With this and Eq. (\ref{MLapprox}), we obtain the $\mathcal{MT}$ symmetric 
dielectric constants $\varepsilon_\parallel=\varepsilon_\perp^*=1.673+0.96593i$ used in this paper.

The width $L_x\sim7.07\mu\mathrm{m}$ of the structure used in this paper
corresponds to 25 periods of multilayers when the lattice constant is $a = 0.2\mu\mathrm{m}$.

\section*{Appendix B: The tilt of the wavefront of the skin mode in the loss-bias case}
\renewcommand{\theequation}{B\arabic{equation}}
\renewcommand{\thefigure}{B\arabic{figure}}
\setcounter{equation}{0}
\setcounter{figure}{0}
In this section, we explain the tilt of the wavefront of the skin mode in the loss-bias case.
First, for a system sandwiched by PECs in the $x$ direction and periodic in the $y$ direction, 
from Eqs. (\ref{Ey})-(\ref{Ex}), each component of the electromagnetic field can be expressed in the following form:
\begin{equation}
  F(x) = e^{iqk_yx}f(x).
\end{equation}
Here, as confirmed in section \ref{MT-setsumei}, in an $\mathcal{MT}$ symmetric medium, $q$ in the Eq. (\ref{q}) is a pure imaginary, 
so we write $q=iq_\mathrm{I}$. Then, 
\begin{equation}
  F(x,y,t)=e^{i(\omega t -k_yy)}e^{-q_\mathrm{I}k_yx}f(x), 
\end{equation}
and the state is localized in the $x$ direction but traveling in the $y$ direction.
In the loss-biased case, on the other hand, $q$ is a complex and can be written as $q=q_\mathrm{R}+iq_\mathrm{I}$, where $q=0.25+0.433i$ 
when $\varepsilon_\parallel=1.673$ and $\varepsilon_\parallel=1.673-1.93186i$. In this case, the electromagnetic field is
\begin{equation}
  F(x,y,t)=e^{i(\omega t +q_\mathrm{R}k_yx-k_yy)}e^{-q_\mathrm{I}k_yx}f(x)
\end{equation}
and thus the wavefront of the skin mode is tilted.
Since it is the component traveling in the azimuthal direction that contributes to OAM, 
in the loss bias case, the tilt of the wavefront may lead to a decrease in OAM.

\section*{Appendix C: Relationship between localization length of skin mode and OAM}
\renewcommand{\theequation}{C\arabic{equation}}
\renewcommand{\thefigure}{C\arabic{figure}}
\setcounter{equation}{0}
\setcounter{figure}{0}
\begin{figure}
  \includegraphics[width=1\columnwidth]{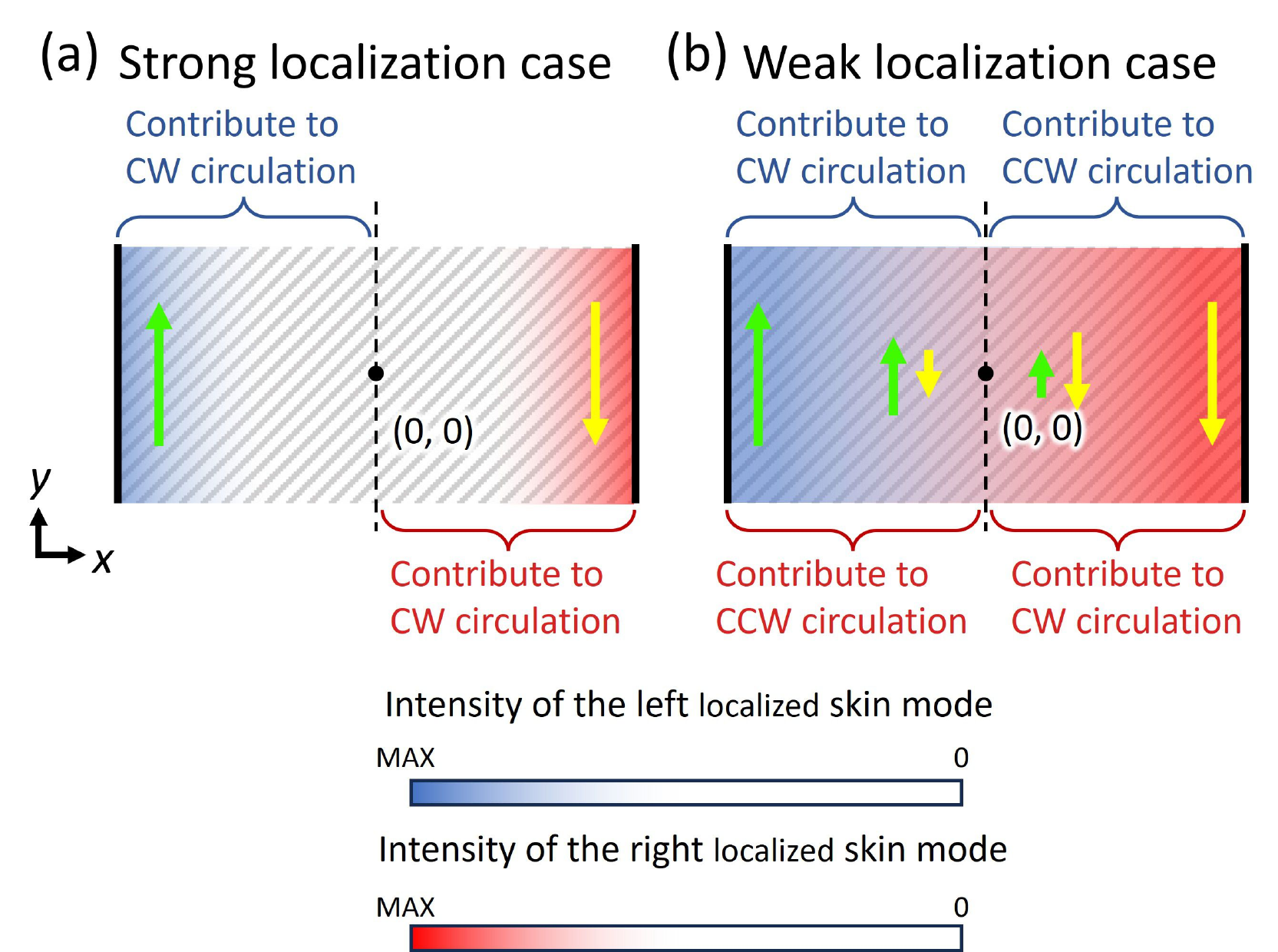}
  \caption{(a)(b) Contribution of skin mode to rotation for short and long localization lengths, respectively.
  Blue and red represent the intensity of the left and right localized skin modes, respectively, 
  and green and yellow arrows represent the pointing vectors of the left and right localized skin modes, respectively.}\label{c1}
\end{figure}
In this section, we discuss the relationship between the localization length of the skin mode and OAM.
First, we consider the case where the localization length of the skin mode is sufficiently short(Fig. \ref{c1}). 
In this case, the skin mode propagation contributes only to unidirectional circulation.
However, as shown in Fig. \ref{c1}, when the localization length is long, the tail of the skin mode will contribute to rotation in the opposite direction.
This would result in a smaller OAM than in the case of shorter localization lengths.

From Eqs. (\ref{Ey})-(\ref{Ex}), $q$ in Eq. (\ref{q}) is one of the quantities that determine the localization length, and when compared for skin modes with the same $k_y$, 
the larger $|\mathrm{Im}(q)|$ is, the shorter the localization length becomes.
Here, in the $\mathcal{MT}$ symmetric medium used in section \ref{MT-circulating}, $|\mathrm{Im}(q)|=0.5774$, and in the loss-biased medium used in section \ref{loss-circulating}, $|\mathrm{Im}(q)|=0.4473$. 
Thus, when compared for similar order modes, the localization length is longer in the loss-bias case. This may also be one of the reasons for the lower OAM shown in section \ref{loss-circulating}.

\bibliography{IsseiTakeda}

\end{document}